\documentclass[prl,twocolumn,amsmath,amssymb]{revtex4}
\usepackage{graphicx}
\usepackage{dcolumn}
\usepackage{bm}
\usepackage{epsfig}
\usepackage{amsmath}
\usepackage{color}
\newcommand{\gam} {\gamma}
\newcommand{\kap} {\kappa}
\newcommand{\wat} {\widetilde{{\omega}}_{a}}
\newcommand{\wct} {\widetilde{{\omega}}_{c}}
\newcommand{\wpt} {\widetilde{{\omega}}_{1+}}
\newcommand{\wmt} {\widetilde{{\omega}}_{1-}}

\newcommand{\Lsig} {L_{\sigma}}

\newcommand{\La} {L_{a}}

\newcommand{\aop} {a}

\newcommand{\caop} {a^{\dag}}

\newcommand{\beq} {\begin{equation}}
\newcommand{\eeq} {\end{equation}}
\newcommand{\bseq} {\begin{subequations}}
\newcommand{\eseq} {\end{subequations}}
\newcommand{\beqz} {\setlength{\mathindent}{0cm}\begin{equation}}
\newcommand{\eeqz} {\end{equation}}
\newcommand{\ber} {\begin{eqnarray}}
\newcommand{\eer} {\end{eqnarray}}
\newcommand{\bers} {\begin{eqnarray*}}
\newcommand{\eers} {\end{eqnarray*}}

\begin{document}

\title{Observation of squeezed light from one atom excited with two photons\footnote{Publication reference: Nature {\bf 474}, 623-626 (2011),

\url{www.nature.com/doifinder/10.1038/nature10170}}}
\author{A. Ourjoumtsev$^{1,2}$}
\author{A. Kubanek$^1$}
\author{M. Koch$^1$}
\author{C. Sames$^1$}
\author{P.W.H. Pinkse$^{1}$\footnote{Present address: MESA+ Institute for Nanotechnology, University of Twente, P.O. Box 217, 7500 AE Enschede, The Netherlands}
}
\author{G. Rempe$^1$}
\author{K. Murr$^1$}
\affiliation{$^1$Max-Planck-Institut f\"ur Quantenoptik,
Hans-Kopfermann-Str.~1, D-85748 Garching, Germany\\
$^2$Laboratoire Charles Fabry de l'Institut d'Optique, CNRS UMR 8501, Universit\'e Paris Sud XI, F-91127 Palaiseau, France}

\date{\today}

\maketitle

\textbf{
Single quantum emitters like atoms are well-known as non-classical light sources which can produce photons one by one at given times\cite{Grangier04}, with reduced intensity noise. However, the light field emitted by a single atom can exhibit much richer dynamics. A prominent example is the predicted\cite{Walls81}   ability for a single atom to produce quadrature-squeezed light\cite{Drummond04}, with sub-shot-noise amplitude or phase fluctuations. It has long been foreseen, though, that such squeezing would be ``at least an order of magnitude more difficult'' to observe than the emission of single photons\cite{Mandel82}. Squeezed beams have been generated using macroscopic and mesoscopic media down to a few tens of atoms\cite{Foster00}, but despite experimental efforts\cite{Hoffges87,Gerber09,Stobinska10}, single-atom squeezing has so far escaped observation. Here we generate squeezed light with a single atom in a high-finesse optical resonator. The strong coupling of the atom to the cavity field induces a genuine quantum mechanical nonlinearity\cite{Schuster08}, several orders of magnitude larger than for usual macroscopic media\cite{Lambrecht96,McCormick07,Corney08}. This produces observable quadrature squeezing\cite{Meystre82,Carmichael85,Carmichael08} with an excitation beam containing on average only two photons per system lifetime. In sharp contrast to the emission of single photons \cite{Short83}, the squeezed light stems from the quantum coherence of photon pairs emitted from the system\cite{Kubanek08}. The ability of a single atom to induce strong coherent interactions between propagating photons opens up new perspectives for photonic quantum logic with single emitters\cite{Shields07,Deppe08,Bishop09,Hofheinz09,Schneebeli09,Rebic09}.}

Unlike in a standard Kerr medium, our squeezing does not result from a simple
nonlinear polarization of the medium but from a cavity-enhanced atomic coherence
which exists for weak coherent driving. Consider a two-state atom with ground and
excited states $|g\,\rangle$ and $|e\,\rangle$. In the absence of a resonator, the
amount of squeezing is governed by the atomic coherence,
$\sigma=|g\,\rangle\langle\,e|$, and the excited-state occupation probability. The
latter produces incoherent scattering which destroys the squeezing. Therefore, the
laser intensity must remain low to preserve the atomic and hence the optical
coherence, see Supplementary Information. Under this condition, the optical
squeezing is determined by the fluctuations of the atomic coherence, $\Delta
\sigma^2=\langle(\,\sigma-\langle\,\sigma\,\rangle\,)^2\rangle$, itself given by
$\Delta \sigma^2=-\langle\,\sigma\,\rangle^{2}$ owing to the fermionic character of
a two-state atom, $\sigma^2=0$. Note that this sets an upper bound to the amount of
squeezing which can be obtained, even when all the light scattered by the atom in all directions is observed.

The presence of the cavity introduces two important ingredients as sketched in Fig.
1. First, the cavity mirrors spatially direct the squeezed light towards the
detectors thus eliminating the need to observe the full $4\pi$ solid angle. Second,
the strong coupling to the optical cavity mode makes the energy-level structure of
the atom-cavity system anharmonic and thus allows for two-photon
transitions\cite{Kubanek08} between the system ground state and the second
dressed-state manifold containing two energy quanta\cite{Schuster08},
$|2\pm\,\rangle$, see Fig. 1a. The net result is an amount of squeezing at the
output mirror that is given by $-K\langle\,\sigma\,\rangle^{2}$, where $K$ depends
only on the frequencies and width of the second manifold $|2\pm\,\rangle$, see
Supplementary Information. It follows that for a given excitation and coherence of
the atom, the squeezing using a cavity is scaled by the factor $K$ which can be
large for strong coupling. In particular, for a resonant excitation of the dressed
states $|2\pm\,\rangle$, $K$ increases with the coherent atom-photon coupling rate,
$g$, relative to the total rate of decoherence ($2\kappa$ for the two-photon
coherence and $\gamma$ for the atomic coherence decay),
$|K|\simeq g/(2\kappa+\gamma)$. We emphasize that squeezed light produced by a single atom in free space would be anti-bunched\cite{Kimble77a}. In our case, the light is squeezed and bunched\cite{Kubanek08}.

Our transition scheme is similar to that of a four-wave mixing process\cite{Agarwal87}, but the
underlying physics is radically different as the scheme arises from the strong
coupling of the quantized cavity field with a two-level atom.
Moreover, the non-linear process differs from that in microwave
experiments\cite{Deleglise08} where short unitary evolutions interrupted by
measurements produce non-classical field states while in our case the squeezed light is generated and propagated out of a dissipative resonator under steady-state
driving conditions.

The nonlinearity appears at a single-atom and single-photon level where quantum
fluctuations play a major role\cite{Foster00,Carmichael08} so that its understanding requires a full quantum treatment rather than a simplified linearized
approach. In the experiment, the quadrature operator of the light
field, $X_\theta$, is measured outside the cavity using a homodyne detection with a
controllable phase $\theta$. In this way we measure the time and normally ordered
(symbol ::) autocorrelation of the quadrature fluctuations $\Delta
X_\theta=X_\theta-\langle X_\theta\rangle$,
\begin{equation}
\langle:\Delta X_\theta(\tau)\Delta X_\theta(0):\rangle=
-\frac{1}{2}\Re(K\langle\,\sigma\,\rangle^{2} f(\tau))\ .
\end{equation}
Here $\Re$ denotes the real part and the function $f(\tau)$ describes the dynamics
of the emission process: the photon pairs cascading via the states $|1\pm\,\rangle
$, detuned by $\Delta_{1\pm}=\omega-\omega_{1\pm}$ with respect to the frequency
$\omega$ of the probe laser, create beatnotes decaying according to the linewidths
$\gamma_{1\pm}$ of these states.  Defining their (complex) detunings as
$\widetilde{{\omega}}_{1\pm}=\Delta_{1\pm}+i\gamma_{1\pm}$ yields
\begin{equation}
f(\tau)= \alpha_+ \exp(i\widetilde{{\omega}}_{1+}\tau) +\alpha_-
\exp(i\widetilde{{\omega}}_{1-}\tau) \,\ ,
\end{equation}
where $\alpha_\pm$ depend only on $\widetilde{{\omega}}_{1\pm}$ and sum up to one
($f(0)=1$). Finally, the spectrum of squeezing is obtained by a Fourier transform.

\begin{figure}
\begin{center}
\includegraphics[width=8.5cm]{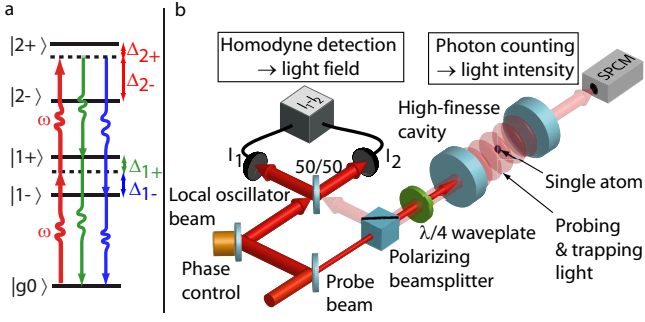}
\caption{  Four-photon process leading to quadrature squeezing for a nearly-resonant
excitation of a single atom strongly coupled to the mode of an optical cavity. The
cavity photon number states $|0\,\rangle, |1\,\rangle , |2\,\rangle ...$ and the
atomic ground and excited states $|g\,\rangle$ and $|e\,\rangle$ combine to form an
anharmonic ladder of dressed states $|n\pm\,\rangle$ sharing $n$ excitations. The
system is excited with laser light of frequency $\omega$. Two laser photons are
absorbed (red arrows) and reemitted down the ladder of states (green and blue
arrows) to produce squeezing. The temporal dynamics of this four-photon process as
well as the spectrum of squeezing is revealed in a homodyne detection scheme, see
sketch in (b): A high-finesse cavity containing a single $^{85}$Rb is excited with a
weak coherent beam. The transmitted photon flux is monitored with a single-photon
counter to control the atom-light coupling. The field properties of the reflected
beam, picked up by an optical circulator (polarizing beam splitter and $\lambda/4$
waveplate), are measured with the balanced homodyne detector. The phase of the
measured quadrature is controlled with a piezoelectric actuator.}
\label{HDSchematicSetup}
\end{center}
\end{figure}
Single $^{85}$Rb atoms are held inside a high-finesse optical cavity by a
red-detuned $785$ nm dipole trap (Fig.1b). A cavity mode, nearly resonant with a
closed atomic transition at $780.24$ nm, is excited with a $P_{in}=8.5$ pW coherent
beam, containing on average $2.0$ photons per cavity decay time (see Supplementary
Information). The effective atom-cavity coupling $g/2\pi=12$ MHz exceeds the atomic
dipole and cavity field decay rates (respectively $\gamma/2\pi=3$ MHz and
$\kappa/2\pi=1.3$ MHz), bringing the system in the strong coupling regime. The
coupling strength is verified by monitoring the transmitted light intensity using a
single-photon counter. The quadratures of the light field reflected from the cavity
are measured with a homodyne detector and sampled with a high-resolution fast
digitizer. After trapping and probing each atom, an additional reference data sample
is acquired with an empty cavity, providing an accurate measurement of the shot
noise level, and the phase of the local oscillator is shifted by $\pm\pi/2$ to
alternate between the $X=X_0$ and $P=X_{\pi/2}$ quadrature measurements. For each
quadrature, we acquire $\approx 3$ s of strong-coupling and $\approx 30$ s of
reference data.

A time-resolved acquisition provides direct access to the quadrature
autocorrelations, revealing the dynamics of the atom-cavity system. For each
quadrature $X_\theta = X$ or $P$ we calculate the time-domain autocorrelations of
the homodyne signal acquired for strongly coupled atoms, and subtract the
autocorrelations of the empty-cavity reference, which leaves a quantity proportional
to $\langle : \Delta X_\theta(\tau) \Delta X_\theta(0):\rangle$. The normalization
factor, obtained by measuring the mean value of the excitation field, perfectly
matches the value expected from the detector's parameters and includes its overall
$55\%$ efficiency without artificially compensating any other experimental
imperfection such as the losses in the atom-cavity system itself.

\begin{figure}
\begin{center}
\includegraphics[width=8.8cm]{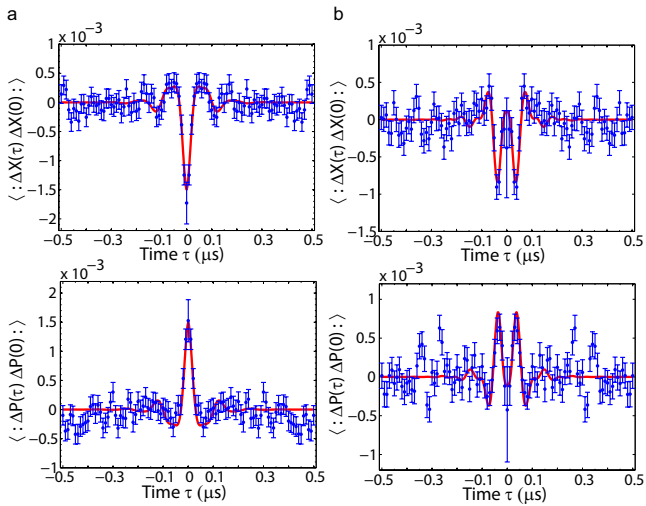}
\caption{Photon beat in the time domain. Autocorrelation functions for the $X$ and
$P$ quadratures. In a, left column, the probe is tuned on the empty cavity
resonance, which favours the transition
$|2+\,\rangle\rightarrow|1+\,\rangle\rightarrow|g0\,\rangle$, with a characteristic
beat via state $|1+\,\rangle$. In b, right column, the probe is close to the
two-photon resonance $|2-\,\rangle$ with an interference between the two possible
paths $|2-\,\rangle\rightarrow|1-\,\rangle\rightarrow|g0\,\rangle$ and
$|2-\,\rangle\rightarrow|1+\,\rangle\rightarrow|g0\,\rangle$. The theoretical curves
are obtained from the analytical model described in the text, taking into account
the extraction efficiency of intracavity photons and the dynamical response of the
homodyne detector. Statistical errorbars correspond to the standard deviations
obtained with $\approx 30\,000$ data samples and $\approx 300\,000$ reference
samples.
}
\label{Autocorrelations}
\end{center}
\end{figure}

Figure 2 presents two homodyne autocorrelation measurements. The first data set,
Fig.2a, is measured with the probe tuned on the empty cavity resonance,
$\omega=\omega_c$, while detuned from the trapped atom by $2\pi\times8$ MHz. The
data present oscillations with a $9$ MHz frequency and a $50$ ns damping time,
characteristic of a beatnote with the closest one-photon dressed state $|1+\rangle$.
The contribution from the state $|1-\rangle$ is negligible, such that
$f(\tau)\propto\exp(i\widetilde{{\omega}}_{1+}\tau)$. The measured autocorrelations
are clearly phase-dependent and the antisymmetry between the $X$ and $P$
quadratures, which translates into an antisymmetry of their noise spectra is a first
sign of nonclassicality. The value at $\tau=0$ corresponds to the difference in the
integrated noise variance between the signal and the reference: the negative value
for the $X$ quadrature confirms the presence of squeezing with one atom. For
comparison, a second set of measurements, shown in Fig.2b, has been performed with
the probe tuned close to the two-photon resonance, $(\omega-\omega_c)/2\pi=-12$ MHz,
while detuned from the atom by $2\pi\times3$ MHz. In this case the probe frequency
is closer to the state $|1-\rangle$ ($\Delta_{1-}/2\pi=9$ MHz) but the transition
through the state $|1+\rangle$ ($\Delta_{1+}/2\pi=-18$ MHz) has a comparable
contribution. With our parameters the two transitions interfere destructively,
making the signal more complex. Moreover, the resonant excitation at the two-photon
transition decreases the phase coherence and therefore, due to a larger contribution
of the incoherent emission, the asymmetry of the $X$ and $P$ autocorrelations is
slightly altered. The theoretical fits obtained from the analytical model above,
including the extraction efficiency of intracavity photons, the dynamical response
of the homodyne detector, and an additional $1$ MHz decoherence rate due to atomic
motion are in very good agreement with the experimental data for both parameter
regimes.

\begin{figure}
\begin{center}
\includegraphics[width=8.8cm]{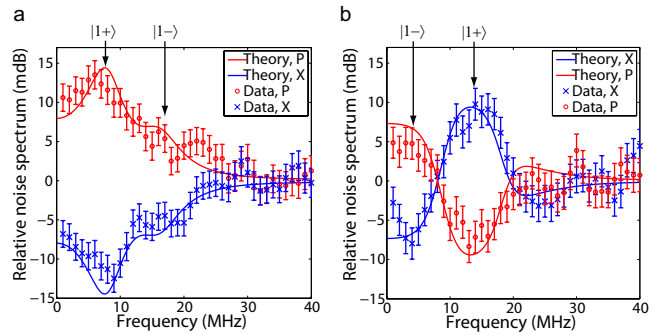}
\caption{Squeezing from one atom. Noise spectra for the $X$ and $P$ quadratures
relative to the reference empty-cavity signal, corrected for the homodyne detector's
efficiency, for the measurement of  Fig. 2a and Fig. 2b. Each curve is an average of
$\approx 30.000$ data and $\approx 300.000$ reference spectra obtained by
Fourier-transforming the homodyne signal over $100$ $\mu$s intervals (see text for
details). The error bars correspond to the standard deviation and account for the
statistical uncertainty and the systematic uncertainty of the shot noise level. The
theoretical curves are obtained from the analytical model in the text, transposed
into the frequency domain.}
\label{Spectra}
\end{center}
\end{figure}
The noise spectra of each quadrature, relative to the shot noise extracted from the
reference data, are determined by a direct Fourier transform of the time-domain
homodyne signal and corrected only for the $55\%$ homodyne detection efficiency.
They are presented in Fig. 3. For $\omega=\omega_c$ (Fig. 3a) at the expected $9$
MHz $\approx |\Delta_{1+}|/2\pi$ frequency, we observe $12\pm 2$ mdB of squeezing on
the $X$ quadrature, with the same amount of antisqueezing on $P$. With a ``perfect''
cavity, single-ended and free from absorption losses, this value would be $5$ times
larger. For $\omega \approx \omega_a$ (Fig. 3b) the negative interference between
the two decay paths results in a cross-over between squeezing and antisqueezing for
each quadrature: the transition through the nearest state $|1-\rangle$ leads to
squeezing on $X$ and antisqueezing on $P$ at low frequencies, whereas the transition
through $|1+\rangle$ leads to the opposite behaviour around $13$ MHz. The spectra
are in remarkable agreement with the analytical model of cavity quantum
electrodynamics for both parameter regimes.

It is instructive to compare the phase-dependent nonlinearity reported here for one
atom to those obtained with macroscopic media. For the measurements with
$\omega=\omega_c$ (Fig. 2a and 3a), where the largest squeezing is observed, the
mean intracavity photon number is only $n=0.033$. Therefore, only a small part of
the impinging beam experiences losses due to atomic spontaneous emission, mirror
coating absorption and cavity transmission, and $\eta=86\%$ of the input power
$P_{in}=8.5$ pW is reflected towards the homodyne detector. For a light beam weakly
squeezed by a third-order non-linear process, we can write an input-output relation
expressing the squeezing in the frequency domain $\langle\Delta \tilde{X}^2
\rangle=-\eta rP_{in}/2$ relative to the shot-noise level of $1/4$, where $r$
determines the nonlinear response of the system. In our case $r\eta=1.6\times10^8$
W$^{-1}$ which exceeds by $7$ orders of magnitude the Kerr nonlinearity of a
standard single-mode optical fibre\cite{Corney08} with the same amount of losses,
and by $4$ orders of magnitude the $\chi^{(3)}$ nonlinearities obtained by four-wave mixing in macroscopic atomic systems with similar
bandwidths\cite{Lambrecht96,McCormick07}. The $\sim10$ mdB squeezing level is of
course very small compared to the $10$ dB achieved in state-of-the-art parametric
upconversion/downconversion experiments\cite{Vahlbruch08}, which remain the best way to generate squeezing as a resource but require several watts of pumping power:
decreasing the latter to a $\sim10$ pW level would bring the squeezing down to
$\sim10^{-9}$ dB. Finally, compared to experiments using atomic beams
\cite{Foster00}, trapping a single atom allows us to keep the non-linearity
constant, as required for most applications.

The squeezing achieved in our setup is limited by the losses in the mirror coatings
and the use of a symmetric cavity with two output ports. Both limitations will be
removed in the near future by means of an asymmetric cavity with a lower loss rate.
Other types of resonators with smaller mode volumes like microtoroids\cite{Dayan08}
or fibre resonators\cite{Gehr10} could allow one to generate squeezed light on an
atom chip. Using artificial atoms like quantum dots in
microcavities\cite{Shields07} would lead to larger and fixed atom-cavity
couplings.
Furthermore, recent theoretical and experimental progress, e.g. \cite{Bozyigit11} indicate that such experiments could soon be transposed to microwave systems using Josephson junctions coupled to strip-line
resonators\cite{Deppe08,Bishop09,Hofheinz09}, where even stronger
non-linearities could be achieved.

\begin{acknowledgments}
Financial support from the Deutsche Forschungsgemeinschaft (Research Unit 635), the European Union (IST project AQUTE) and the Bavarian PhD programme of excellence (QCCC) is gratefully acknowledged.
Correspondence and requests for materials should be addressed to A.O.~(alexei.ourjoumtsev@institutoptique.fr) or K.M.~(karim.murr@gmail.com).
\end{acknowledgments}


\pagebreak
\begin{widetext}
\section{Observation of squeezed light from one atom excited with two photons : Supplementary Information}
\end{widetext}

\section{Theory}

In this section we develop the theoretical description of the squeezing process in our atom-cavity system and outline the derivation of the equations used to interpolate the experimental data. To complement previous approaches based on the pure-state formalism\cite{Carmichael08,Carmichael91}, we make a direct connection between the squeezing of the atomic dipole and the squeezing of the cavity field, and point at a distinction between the contributions of single-photon and two-photon excitation processes.

\subsection{Quadrature squeezing of the cavity field}
We consider a single cavity mode defined by the annihilation and creation operators, $a$ and $a^\dag$ respectively, and obeying the canonical commutation rules $[a,a^\dag]=1$. The quadratures of the intracavity field are defined by
\ber
X_\theta=\frac{1}{2}( e^{-i\theta}\aop+e^{i\theta}\caop ),
\eer
where $\theta$ is an adjustable phase. Defining the fluctuation of $a$ with respect to the steady state value $\langle\, a\,\rangle$ as $\Delta a=a-\langle\, a\,\rangle$, the quadrature variance $\langle\Delta X_\theta^2\rangle=\langle(X_\theta-\langle X_\theta\rangle)^2\rangle$ reads:
\begin{multline}
\langle\Delta X_\theta^2\rangle
=\frac{1}{2}(\Re(e^{-2i\theta}\langle\Delta a^{2}\rangle)+\langle\Delta a^\dagger \Delta a\rangle)+\frac{1}{4},
\end{multline}
which, when normally ordered (symbol $::$), reduces to
\begin{multline}
\langle:\Delta X_\theta^2:\rangle=\frac{1}{2}(\Re(e^{-2i\theta}\langle\Delta a^{2}\rangle)+\langle\Delta a^\dagger \Delta a\rangle)\ .
\end{multline}
For a coherent or a vacuum state, one has $\langle\Delta a^{2}\rangle=\langle\Delta a^\dagger \Delta a\rangle=0$, thus the quadrature variance is equal to the shot noise, $\langle\Delta X_\theta^2\rangle=1/4$, i.e. $\langle:\Delta X_\theta^2:\rangle=0$. The state of the field will be quadrature-squeezed if for some $\theta$ the variance of $X_\theta$ drops below the shot noise, $\langle:\Delta X_\theta^2:\rangle <0$.

The term $\langle\Delta a^\dagger \Delta a\rangle$ is the incoherent part of the spectrum and can never be negative. In order to observe squeezing in a nearly-resonant dissipative system such as ours, this term must remain as small as possible, which can be achieved at weak enough excitation intensities. In this coherent (phase-sensitive) limit it becomes negligible compared to the coherent term $\langle\Delta a^{2}\rangle$ and the quadrature squeezing reduces to
\beq
\label{EqDX0}
\langle : \Delta X_\theta^2 : \rangle
\approx\frac{1}{2}\Re(e^{-2i\theta}\langle\Delta a^{2}\rangle)\ .
\eeq

\subsection{Squeezing generated by one atom in an optical cavity}
To calculate $\langle\Delta a^{2}\rangle$, we consider a two state atom, with a ground state $|g\,\rangle$ and an excited state $|e\,\rangle$, interacting with the mode of the cavity with photon number states $|0\,\rangle, |1\,\rangle , |2\,\rangle ...$. The relevant physical parameters are:
\begin{itemize}
\item{The atom-cavity coupling $g$,}
\item{The cavity field decay rate $\kappa$,}
\item{The atomic dipole decay rate $\gamma$,}
\item{The frequency of the driving laser $\omega$,}
\item{The amplitude of the driving laser $\epsilon$,}
\item{The detuning between the driving beam and the cavity resonance frequency $\Delta_c=\omega-\omega_c$,}
\item{The detuning between the driving beam and the atomic resonance frequency $\Delta_a=\omega-\omega_a$.}
\end{itemize}
The evolution of the system's density matrix $\rho$ is given by a master equation
\beq
\frac{d\rho}{dt}=\mathcal{L}\rho=-\frac{i}{\hbar}[H_{JC}+H_P,\rho]
+\kappa\La\rho+\gamma\Lsig\rho\ .
\eeq
In the rotating wave approximation and in the interaction picture with respect to the laser frequency, the Jaynes-Cummings hamiltonian is
\ber
H_{JC}/\hbar=-\Delta_c a^\dag a -\Delta_a \sigma^\dag \sigma +g (a^\dag \sigma + a\sigma^\dag),
\eer
where $\sigma=|g\rangle\langle e|$ and $\sigma^\dag=|e\rangle\langle g|$ are, respectively, the lowering and raising operators. The terms
\ber
\kap\La\rho&=&\kap(2a\rho a^\dag-\rho a^\dag a-a^\dag a\rho)\\
\gam\Lsig\rho&=&\gam(2\sigma\rho\sigma^\dag-\rho \sigma^\dag \sigma-\sigma^\dag \sigma\rho),
\eer
respectively describe the losses through the cavity mirrors and spontaneous emission from the atom into the free space modes.
The laser light incident on the input mirror results in a coherent excitation of the cavity mode:
\ber
H_{P}/\hbar=\epsilon (a^\dag + a)\ .
\eer

The time evolution of the expectation value of a time-independent operator
$\mathcal O$ is computed as $d\langle \mathcal O\rangle/dt=Tr(\mathcal O\mathcal{L}\rho)$. Defining complex detunings as $\wct=\Delta_c+i\kappa, \wat=\Delta_a+i\gamma$, we have
\ber
&&\frac{d}{dt}\langle a\rangle=i(\wct \langle a\rangle-g\langle\sigma\rangle-\epsilon)\\
&&\frac{d}{dt}\langle \sigma\rangle=i(\wat \langle \sigma \rangle+g\langle a\sigma_z\rangle)\\
&&\frac{d}{dt}\langle a^2\rangle=2i(\wct \langle a^2\rangle -g\langle a \sigma\rangle-\epsilon\langle a\rangle)\\
&&\frac{d}{dt}\langle a\sigma\rangle=i((\wat+\wct) \langle a\sigma \rangle+g\langle a^2\sigma_z\rangle-\epsilon\langle \sigma \rangle)\;\;\;\;\;\;\;
\eer
where $\sigma_z=|e\rangle\langle e|-|g\rangle\langle g|$ is the population inversion of the atom.

We operate in the weak excitation regime, where the population of the excited atomic state $|e\rangle$ is negligible. In this case $\langle \sigma_z\rangle\approx -1$, $\langle a\sigma_z\rangle\approx -\langle a\rangle$, $\langle a^2\sigma_z\rangle\approx -\langle a^2\rangle$, and the equations above provide the steady-state solutions
\ber
\langle a\rangle&=&\frac{\epsilon \wat}{\wat\wct-g^2},\\
\langle \sigma\rangle&=&\frac{\epsilon g}{\wat\wct-g^2},\\
\langle a^2\rangle&=&\frac{\epsilon^2(\wat(\wat+
\wct)+g^2)}{(\wct(\wat+\wct)-g^2)(\wat\wct-g^2)},\\
\langle a\sigma\rangle&=&\frac{\epsilon^2 g(\wat+\wct)}{(\wct(\wat+\wct)-g^2)(\wat\wct-g^2)},
\eer
which yields for the fluctuations:
\ber
\langle \Delta a^2\rangle&=&\frac{-\epsilon^2g^4}{(\wct(\wat+\wct)-g^2)
(\wat\wct-g^2)^2}\;\;\;\;\;\label{EqDeltAA}\\
\langle \Delta a\Delta \sigma\rangle&=&\frac{-\epsilon^2 g^3\wct}{(\wct(\wat+\wct)-g^2)(\wat\wct-g^2)^2}\label{EqDeltAS}
\eer
Notice that, under the assumption of weak excitation, the field amplitude $\langle a\rangle$ and the atomic coherence $\langle\sigma\rangle$ scale as $\epsilon$, whereas $\langle a^2\rangle$ and $\langle a\sigma\rangle$ scale as $\epsilon^2$. More involved calculations\cite{Carmichael08,Carmichael91} would show that $\langle\Delta a^\dagger \Delta a\rangle$ scales as $\epsilon^4$. This is nothing else than saying that the spectrum is coherent at weak excitation, $\langle a^\dag a\rangle\approx|\langle a \rangle|^2\propto\epsilon^2$, thereby justifying the use of Eq.\ref{EqDeltAA} to compute Eq.\ref{EqDX0}.

We now define the complex detunings of the dressed states $|n\pm\rangle$ as
\ber
\nonumber\widetilde{{\omega}}_{n\pm}&=&(n-1)\wct+\frac{1}{2}(\wct+\wat)\\
&&\mp\frac{1}{2}\sqrt{4ng^2+(\wct-\wat)^2}\ .
\eer
This allows us to simplify
\ber
\label{EqSig}
\langle \sigma\rangle&=&\frac{\epsilon g}{\widetilde{{\omega}}_{1+}
\widetilde{{\omega}}_{1-}},\\
\langle \Delta a^2\rangle&=&K(-\langle\,\sigma\,\rangle^{2}),
\eer
where the constant $K$ is simply
\beq
\label{EqK}
K=\frac{2g^{2}}{\widetilde{{\omega}}_{2+}\widetilde{{\omega}}_{2-}}\ ,
\eeq
with the result in the body of the paper,
\beq
\label{EqDeltXX}
\langle:\Delta X_\theta^2:\rangle=-\frac{1}{2}\Re(e^{-2i\theta}K\langle\,\sigma\,\rangle^{2})\ .
\eeq

\subsection{Physical interpretation of the squeezing mechanism}
Equation \ref{EqDeltXX} shows that the squeezing of the optical field in the cavity is directly related to the squeezing of the atomic coherence. Due to the two-level structure of the atom, the elementary property $\sigma^2=0$ directly gives $\langle(\,\sigma-\langle\,\sigma\,\rangle\,)^2\rangle=
-\langle\,\sigma\,\rangle^{2}$. In the dressed-state picture, $\langle\,\sigma\,\rangle$ depends only on the normal modes $|1\pm\rangle$, see Eq.\ref{EqSig}, thus highlighting the one excitation processes. The constant $K$, in contrast, connects the squeezing of the cavity field to the squeezing of the atomic coherence and demonstrates the importance of the two-photon states $|2\pm\rangle$, as obvious from its definition Eq.\ref{EqK}. The observed squeezing is therefore fundamentally dependent on the anharmonic structure of the atom-cavity system. The underlying physical mechanism can be summarized as such: At low atomic excitation, the dynamics of the internal state of the atom are largely coherent, which due to the quantized nature of the atom (two-level structure) directly translates into polarization squeezing, $-\langle\,\sigma\,\rangle^{2}$, and couples back to the cavity mode to create quadrature squeezing, $-K\langle\,\sigma\,\rangle^{2}$.

\subsection{Time-domain evolution and squeezing spectrum}
The quantum regression theorem provides a set of linear differential equations for the two-time correlations:
\ber
\nonumber\frac{d}{d\tau}\langle \Delta a(\tau)\Delta a(0)\rangle
&=&i(\wct \langle\Delta a(\tau)\Delta a(0)\rangle\\
&&-g\langle\Delta \sigma(\tau)\Delta a(0)\rangle),\\
\nonumber\frac{d}{d\tau}\langle\Delta \sigma(\tau)\Delta a(0)\rangle
&=&i(\wat \langle\Delta \sigma(\tau)\Delta a(0)\rangle\\
&&-g\langle\Delta a(\tau)\Delta a(0)\rangle).
\eer
Solved with the initial conditions given by Eqs. \ref{EqDeltAA} and \ref{EqDeltAS}, it yields for $\tau>0$
\beq
\langle:\Delta X_\theta(\tau)\Delta X_\theta(0):\rangle=-\frac{1}{2}\Re (e^{-2i\theta}K\langle\sigma\rangle^2 f(\tau))\ ,\label{EqDeltXtX0}
\eeq
where the regression of the fluctuations is given by
\ber
f(\tau)&=&
\frac{\widetilde{{\omega}}_{1+}}{\widetilde{{\omega}}_{1+}-\widetilde{{\omega}}_{1-}}
\exp(i\widetilde{{\omega}}_{1-}\tau)\\
&&- \frac{\widetilde{{\omega}}_{1-}}{\widetilde{{\omega}}_{1+}-\widetilde{{\omega}}_{1-}} \exp(i\widetilde{{\omega}}_{1+}\tau).\;\;\;
\eer
The squeezing spectrum measured outside the cavity, normalized to the shot noise, can be obtained by a simple Fourier transform of the autocorrelations. Taking into account the overall detection efficiency $\eta$ and the fact that the fluctuations of the field leaking outside the cavity can be related to those inside by $\langle\Delta a_{out}^2\rangle=2\kappa\langle\Delta a^2\rangle$, it reads
\begin{multline}
S_\theta(\Omega)=1+\eta2\kappa\int_{0}^\infty d\tau\cos(\Omega\tau)\langle:\Delta X_\theta(\tau)\Delta X_\theta(0):\rangle\;\;\;\;\;\;\;\;\;\;\\
=1+8\eta\kappa\Re\left\{e^{-2i\theta}\langle\Delta a^2\rangle\frac{i\wpt\wmt}{\wpt-\wmt}\right.\\
\left.\times\left(\frac{1}{\Omega^2-\wmt^2}-\frac{1}{\Omega^2-\wpt^2}\right) \right\},
\end{multline}
sum of two Lorentzians with frequencies and widths defined by the normal modes.

\subsection{Interpretation in the dressed-state picture}
The physical description can now be transposed into the dressed-state picture to determine the transition mechanism shown on Fig. 1 in the main paper body. Two pump photons make a near-resonant excitation of the second doublet of dressed states, $|2\pm\rangle$, which decay via the normal modes $|1\pm\rangle$. The time dependence of the measured signal, determined by $f(\tau)$, corresponds to the beatnote of these photons with the local oscillator which has the same frequency $\omega$ as the probe. It is the sum of two terms oscillating at the detuning frequencies $\omega-\omega_{1\pm}$ of the normal modes and damping according to their linewidths. For each of the two possible transitions, the two emitted photons appear in opposite sidebands with respect to the local oscillator and, due to the coherence of the emission process, this results in quadrature squeezing at the corresponding frequency.

\section{Experimental setup}
The atom-cavity system is based on a $123$\,$\mu$m - long Fabry-Perot optical resonator formed by two identical mirrors (transmission $2.8$\,ppm, absorption losses $4.0$\,ppm, radius of curvature $200$\,mm, diameter $7.75$\,mm), reaching a finesse of $\approx 470.000$. A $785$\,nm laser beam, resonant with a TEM$_{00}$ cavity mode, is used to stabilize the cavity length and to form a red-detuned dipole trap for single $^{85}$Rb atoms, injected into the cavity by an atomic fountain. A $780.24$ nm probing beam excites another TEM$_{00}$ cavity mode, nearly resonant with the $(5^2S_{1/2}, \; F=3, \;m_F=3 \rightarrow 5^2P_{3/2}, \; F=4, \; m_F=4)$ atomic transition. The probing power impinging on the cavity is set to $8.5$\,pW, which corresponds to $2.0$ photons per relevant temporal mode defined by the cavity decay time of $60$\,ns. The effective atom-light coupling $g/2\pi=12$\,MHz exceeds the atomic dipole and cavity field decay rates (respectively $\gamma/2\pi=3$\,MHz and $\kappa/2\pi=1.3$\,MHz) and brings the system into the strong-coupling regime. The quality of the coupling, reduced compared to the maximal value $g_{max}/2\pi=16$\,MHz by the motion of the atom inside the trap, is deduced from the cavity transmission measured with a single photon counter: strongly coupled atoms drive the cavity off resonance and decrease the transmitted photon flux by a factor of $50$ \cite{Schuster08}. The light exiting the cavity through the input mirror is picked up by an optical circulator and reflected towards a balanced homodyne detector (Thorlabs PDB120A). The phase of the local oscillator, controlled by two mirrors mounted on piezoelectric actuators, is locked using an auxiliary $772$\,nm frequency-stabilized beam far off-resonant with respect to any cavity mode and hence directly reflected by the input cavity mirror. The homodyne signal is digitized with a 14-bit resolution at a $100$\,MHz rate. Within the overall $40$\,MHz bandwidth of the homodyne detection system, limited by the anti-alias filter of the digitizer, the $800$\,$\mu$W local oscillator power yields a shot noise level $\geq 9$\,dB above the electronic noise.

\section{Experimental sequence}
The experimental sequence is divided in three parts: preparation, measurement, and control. During the preparation time ($\approx 3$\,s), atoms are loaded into a magneto-optical trap located $25$\,cm below the cavity, optically cooled down to $\approx 5$ $\mu$K, and launched towards the cavity which they reach at nearly zero velocity. A sharp drop in the cavity transmission heralds the arrival of a well-coupled atom, trapped by raising the dipole potential from $0.2$ to $0.9$\,mK. This triggers a $20$ ms measurement sequence, which depends on the probing frequency. When the probe beam is resonant with the cavity ($\omega=\omega_c$), a cavity cooling mechanism leads to good atomic storage times and the system is probed continuously with a $P_{in}=8.5$\,pW incoming power. When the probe is tuned close to the atomic resonance ($\omega \approx \omega_a$) no cooling occurs: in this case the measurement sequence is divided into alternating $200$\,$\mu$s probing and cooling intervals (the probe power and frequency being set to $P_{in}=8.5$\,pW, $\omega \approx \omega_a$ for probing and $P_{in}=1.7$\,pW, $\omega = \omega_c$ for cooling). After each measurement, the last part of the sequence ($\approx 1$\,s) is used to verify the powers of the trapping and cooling beams, as well as the power and the phase of the local oscillator, and to shift this phase by $\pm\pi/2$ to alternate between the $X$ and $P$ quadrature measurements.

The extremely low signal level combined with short atomic storage times puts drastic requirements on the experimental stability and technical noise suppression. This is achieved by actively controlling all experimental parameters and by optimizing the data acquisition procedure.  During the $20$ ms measurement sequence, the atom remains strongly coupled for $1.6$\,ms on average before it leaves the cavity. The remaining data, acquired when the cavity is empty, is used as a reference. This nearly simultaneous signal and reference acquisition leads to an excellent cancelation of technical noises and slow thermal drifts. For each probing frequency we repeat the measurement sequence during $\sim 100$\,h to acquire $\approx 3$\,s of strong-coupling and $\approx 30$\,s of reference data for each quadrature.

\section{Data analysis}
For each trapped atom, we divide the $20$\,ms data sample into $200\,\mu$\,s intervals and, using the measured cavity transmission, we select those where the atom was strongly coupled (transmission $T\leq 0.04\,T_{max}$), and those where the cavity was empty ($T\geq 0.7\,T_{max}$). For each interval, we calculate the autocorrelations with the maximal $10$\,ns time resolution allowed by the digitizer's speed, we average the calculated functions over all qualified intervals with a given parameter set, and we subtract the mean value at large time delays ($t>1$\,$\mu$s) where signals become uncorrelated.

\begin{figure}
\begin{center}
\includegraphics[width=8cm]{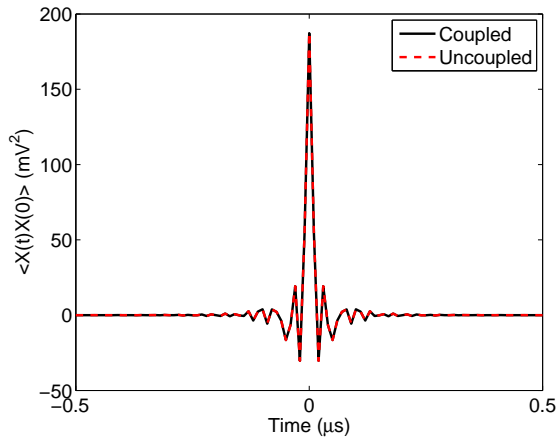}
\caption{``Raw'' homodyne autocorrelations, comparing the signal (strongly coupled atom) and reference (uncoupled atom) data for a given parameter set ($X$ quadrature, probe frequency $\omega=\omega_c$, for other parameters the curves are very similar). The visible oscillations correspond to the impulse response of the detector, i.e. to the autocorrelations of the shot noise measured with the finite detection bandwidth.}
\label{RawAutocorr}
\end{center}
\end{figure}

Figure \ref{RawAutocorr} shows that the obtained ``raw'' autocorrelations of the strong-coupling signal are practically superimposed with those of the empty-cavity reference. The oscillations visible at this scale correspond only to the impulse response of the detector, i.e. to the autocorrelations of the shot noise measured with a finite bandwidth. Physically relevant signals, $\approx 10^4$ times smaller, become only accessible by calculating the difference of these curves, where these large oscillations cancel out, leaving a quantity proportional to ${\langle :\Delta X_\theta(\tau) \Delta X_\theta(0):\rangle}$. In order to properly normalize it, we unlock the cavity and measure the offset on the $X$ quadrature: all of the light is then reflected towards the homodyne detector, which corresponds to measuring a coherent state $|\alpha\rangle$ with $\langle X\rangle=\alpha=\sqrt{2.0}$ and $\langle P\rangle=0$. The corresponding normalization factor perfectly matches the value expected from the detector's parameters and includes its overall efficiency $\eta_d=0.55$, without artificially compensating any other experimental imperfection such as the losses in the atom-cavity system itself.

An accurate subtraction between the signal and the reference requires an extremely stable shot noise level, determined by the local oscillator's power. This power is actively stabilized during the preparation and the control phases of the measurement sequence, but the stabilization system is too slow to follow the fast switching between the $200$\,$\mu$s probing and cooling intervals, and during the $20$\,ms measurement time the power is controlled by a sample-and-hold circuit. This circuit presents a small systematic drift, and the shot noise level changes by $\approx10$\,dBm (i.e. $\approx 0.2\%$) during these $20$\,ms. Since the strong coupling data corresponds to the beginning, and the reference data to the end of the acquisition, this variation appears on the differential autocorrelations, although it is too small to be visible on Fig. \ref{RawAutocorr}. To measure and compensate for this drift, for each parameter set we select the sequences where the atom was not trapped at all (triggering the acquisition but leaving the trap within the first $100$\,$\mu$s). In this case the variance of the homodyne signal corresponds only to the shot noise. By averaging over all such sequences, we obtain the evolution of the shot noise during the $20$\,ms acquisition. We then determine the average time interval when the atoms stayed strongly coupled (first $1.6$\,ms) and the one when the cavity was empty (last $16$\,ms of the $20$\,ms acquisition) and, for each case, calculate the average shot noise variance. Rescaling the reference data by their ratio $\zeta$ (typically $0.998$) compensates for the drift and allows to completely cancel out the contribution of the shot noise on the differential autocorrelations: dominant on Fig. \ref{RawAutocorr}, it is absent from Fig. 2 in the main body of the paper.

Although the agreement between the measured data and the theoretical autocorrelation function $\langle:\Delta X_\theta(\tau)\Delta X_\theta(0):\rangle$ (Eq. \ref{EqDeltXtX0}) is satisfactory, it can be further improved by taking into account the percussional response of the detector $D(\tau)$, obtained by renormalizing to 1 the integrated shot noise autocorrelations on Fig. \ref{RawAutocorr}. The theoretical curves on Fig. 2 in the main paper body correspond convolution of the theoretical autocorrelations $\langle:\Delta X_\theta(\tau)\Delta X_\theta(0):\rangle$ with $D(\tau)$.

\begin{figure}
\begin{center}
\includegraphics[width=8cm]{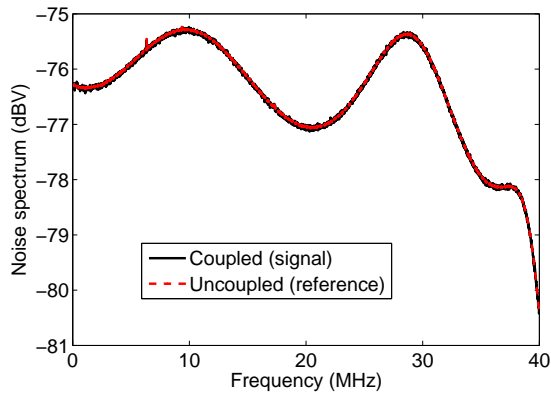}
\caption{``Raw'' signal and reference homodyne noise spectra ($X$ quadrature, $\omega=\omega_c$, for other parameters the curves are very similar). The visible features correspond to the frequency dependence of the detection gain, defined by the anti-alias filter (steeply rolling off beyond $40$\,MHz), and to a small ($200$\,mdB) and narrow ($<10$\,kHz) residual technical noise peak at $6.3$\,MHz with an equal height on both curves. The DC offset peak ($-38.9$\,dBV) is cut off.}
\label{RawSpectra}
\end{center}
\end{figure}

We calculate the noise spectra with a $10$\,kHz resolution by a direct Fourier transform of the time-domain homodyne signal divided into $100$\,$\mu$s intervals. Figure \ref{RawSpectra} shows that the ``raw'' spectra of the signal and the reference are, like the autocorrelations, completely superimposed, the reference spectrum being less noisy due to better statistics. The only visible features are the smooth variations of the shot noise level due to a non-flat response of the digitizer's anti-alias filter, and a small ($200$\,mdB) and narrow ($<10$\,kHz) residual technical noise peak at $6.3$\,MHz with an equal height on both curves. We rescale the reference spectrum by the same factor $\zeta$ as the autocorrelations to compensate for the small shot noise drift, and calculate the difference between the two spectra in logarithmic scale. Like for the autocorrelations, this cancels out the residual technical noise, and brings the noise level for frequencies above $30$\,MHz within $1$\,mdB from $0$: as can be expected from the parameters of the atom-cavity system, no spectral features appear at those frequencies, and the noise level of the signal is equal to the shot noise level of the reference. Since the narrow $10$\,kHz resolution is only necessary for efficient technical noise cancelation and is otherwise excessive for a system where relevant decay times remain below $0.1$\,$\mu$s, we average out narrow-band spectral fluctuations by convolution with a $1$\,MHz Lorentzian filter, which amounts to assuming that signals separated by $\tau>0.16$\,$\mu$s are uncorrelated. Finally, we correct the obtained spectra by the independently measured $55\%$ homodyne detection efficiency, to obtain the curves presented on Fig.\,3 in the paper body. We verified that the same spectra can be obtained by Fourier-transforming the differential autocorrelation functions and rescaling the result by the cavity decay rate and the overall detection efficiency including cavity losses.

\end{document}